\documentclass{elsart}
\usepackage[square, comma]{natbib}
\usepackage{graphics}
\usepackage{amsmath}
\usepackage[psamsfonts]{amssymb}
\usepackage[figuresright]{rotating}
\journal{Physics Letters B}
\volume{611}
\firstpage{223}
\lastpage{230}

\def\Journal#1#2#3#4{{#1} {\bf #2} (#3) #4}

\def\PLB{{\em Phys. Lett.}  B}

\def\PRD{{\em Phys. Rev.} D}
\def\ZPHY{{\em Z. Phys.} C}
\def\PRP{\em Phys. Rep.}
\def\Chione{\tilde{\chi}^{\pm}_{1}}
\def\Chizero{\tilde{\chi}^{0}_{1}}
\def\Chitwo{\tilde{\chi}^{0}_{2}}
\def\Stau{\tilde{\tau}^{\pm}_{1}}

\begin{document}
\runauthor{Kato, Fujii, Kamon, Khotilovich and Nojiri}
\begin{frontmatter}
\title{Mass and Cross-section Measurements of Chargino at 
Linear Colliders in Large \mbox{\boldmath $\tan\beta$} Case}
\author[Kinki]{Yukihiro Kato\corauthref{cor}},
\corauth[cor]{Corresponding author. \\
 addresses: Department of Physics, Kinki University, 3-4-1 Kowakae, 
Higashi-Osaka, Osaka, 577-8502, Japan}
\ead{katoy@hep.kindai.ac.jp}
\author[KEK]{Keisuke Fujii},
\author[Texas]{Teruki Kamon},
\author[Texas]{Vadim Khotilovich},
\author[Kyoto]{Mihoko M. Nojiri}

\address[Kinki]{Department of Physics, Kinki University, Osaka, Japan}
\address[KEK]{High Energy Accelerator Research Organization, Ibaraki, Japan}
\address[Texas]{Department of Physics, Texas A\&M University, 
College Station, TX, U.S.A.}
\address[Kyoto]{Yukawa Institute for Theoretical Physics, 
Kyoto University, Kyoto, Japan}


\begin{abstract}
The lighter chargino pair production ($e^+e^- \to \tilde{\chi}^{+}_{1}
\tilde{\chi}^{-}_{1}$) is one of the key processes for determination
of supersymmetric parameters at a linear collider.
If $\tan\beta$\ is a large value, the lighter stau ($\Stau$) might be
lighter than the $\Chione$, while the other sleptons stay heavier.
This case leads to a cascade decay, $\Chione \rightarrow 
\Stau \nu_{\tau}$\ followed by $\Stau \rightarrow \Chizero \tau^{\pm}$,
where $\Chizero$\ is the lightest supersymmetric particle.
This paper addresses to what extent this cascade decay 
may affect the measurements of
the chargino mass and its production cross section.
\end{abstract}
\end{frontmatter}

\newpage

\section{Introduction}
Supersymmetry (SUSY) is one of the most promising scenarios of
physics beyond the Standard Model (SM) and predicts new particles
observable at the present or next generation of colliders.
It solves the naturalness problem by taming the otherwise quadratically 
divergent quantum correction to the Higgs boson mass,
makes the three gauge forces unify, 
and opens up a way to ultimate unification of
all the fundamental interactions including gravity \cite{mssm}. 
If SUSY is the case, we will be able to probe physics at the
unification scale by studying its breaking pattern 
through various measurements of SUSY particles
some of which will have been discovered at the Tevatron or the LHC.
One of primary goals for an International Linear Collider (ILC) is hence 
determination of SUSY or SUSY-breaking parameters 
through precision measurements of masses and couplings of
those SUSY particles.
The pair production of lighter charginos 
($e^{+}e^{-} \rightarrow \tilde{\chi}^{+}_{1} \tilde{\chi}^{-}_{1}$)
is one of the most important processes in this respect 
since the $\Chione$\ is relatively light 
in most of the parameter space
and hence expected to be studied in the early stage of the ILC project.
Measurements of its mass and couplings at the ILC 
have not only been discussed but, 
for some typical sets of SUSY(-breaking) parameters, also
simulated in detail~\cite{Ref:susysim}. 

All of the cases studied in the past, however, assume that
the lighter chargino would decay directly into a final state
containing the lightest supersymmetric particle (LSP), 
which is the lightest neutralino ($\Chizero$) 
in most of the SUSY parameter space.
At large $\tan\beta$\ values, however,
$\tilde{\tau}_{L}$-$\tilde{\tau}_{R}$ mixing
might generate a large mass splitting between
$\tilde{\tau}_1$\
and the other sleptons
($\tilde{\ell} =\ \tilde{e}$, $\tilde{\mu}$, $\tilde{\tau}_2$, 
$\tilde{\nu}$),
yielding a particular mass hierarchy of
$M_{\tilde{\ell}} >  M_{\tilde{\chi}^{\pm}_{1}} > M_{\tilde{\tau}^{\pm}_{1}}
 > M_{\tilde{\chi}^{0}_{1}}$.
If $\tilde{\chi}^{\pm}_{1}$ is gaugino-like, this means that
the $\Chione$ will decay at 100\% of the times in the cascade mode: 
$\Chione \rightarrow \Stau \nu_{\tau}$\ 
followed by 
$\Stau \rightarrow \Chizero \tau^{\pm}$.
Consequently, the experimental signature of the chargino
pair production will be an acoplanar $\tau$ pair with
significant missing energy carried away not only by
the LSP's from the $\Stau$ decays but also by
the final-state neutrinos from the $\Chione$\ and $\tau$ decays.
This would make it significantly more difficult to 
determine the chargino mass. 

This paper describes results from our simulation studies 
of the pair production of
the lighter charginos and their subsequent cascade decays:
$e^+ e^- \rightarrow \tilde{\chi}^{+}_{1} \tilde{\chi}^{-}_{1} \rightarrow 
\tilde{\tau}^{+}_{1}\nu_{\tau} + \tilde{\tau}^{-}_{1}\overline{\nu}_{\tau}$\
in a particular scenario of the Minimal Supersymmetric
Standard Model (MSSM),
where $M_{\tilde{\tau}^{\pm}_{1}} = 152.7$\ GeV/$c^{2}$\ and 
$M_{\Chizero} = 86.4$\ GeV/$c^{2}$\ as in 
Refs.~\cite{stau,stau2}. 
It should be noted that signals from the $\tilde{\tau}^{\pm}$-$\Chione$\ 
co-annihilation region
are recently of great interest. 
The region is characterized by a difference in the $\tilde{\tau}^{\pm}$\  
mass and the lightest neutralino mass of about 5-15 GeV/$c^{2}$.
This small mass difference allows the $\tilde{\tau}^{\pm}$\  to 
co-annihilate in the 
early universe along with the neutralinos in order to produce the current 
amount of cold dark matter density of the universe measured 
by WMAP~\cite{wmap}.
However, the experimental signature is very challenging.
 We will not address this signature, because it requires to detect 
electron/positron from two-photon process in the forward region. Thus it
requires a different optimization of the event selection 
criteria~\cite{coann1, coann2, coann3}.

\section{Monte Carlo Simulation}
In simulating the signal and various background events,
we use the HELAS library~\cite{HELAS}
to calculate the helicity amplitudes,
the BASES/SPRING package~\cite{BASES} to 
generate the final-state partons,
and TAUOLA~\cite{TAUOLA} to decay
the final-state $\tau$ leptons if any,
so as to take into account possible spin correlations of
intermediate heavy partons, 
effects of beamstrahlung and subsequent 
beam-energy spread~\cite{bespd},
and finite $\tau$ polarizations.
The generated events are then processed through a fast simulator of
a detector model~\cite{JLC},
which includes a central tracker, 
electromagnetic and hadron calorimeters, 
and muon drift chambers. 
This simulator smears charged-track parameters in the central
tracker with parameter correlation properly taken into account, 
and simulates calorimeter signals as from individual segments.
Although $\tau$\ polarization varies in general 
with the SUSY parameters
and would thus influence 
the property of the final states, 
we fix the $\tau$\ polarization for the signal process 
at the nominal value of 0.026 for simplicity. 

For SUSY events, we choose our reference SUSY point to have  
$\tan\beta = 50$, $\mu = 400$\ GeV,
m$_{0} = 200$\ GeV, and M$_{2}$ = 180.0 GeV,
where $\mu$, m$_{0}$, and M$_{2}$\ are 
the higgsino mixing mass, 
universal scalar mass at GUT scale, 
and SU(2)$_{\rm L}$\ gaugino mass parameters, respectively. 
We take M$_{1}$\ so that it satisfies GUT relation M$_{1} \cong 0.5$M$_{2}$. 
It corresponds 
to $M_{\tilde{\chi}^{\pm}_{1}} =171.0$\ GeV/$c^{2}$\ ($M_{\Chione}^{\rm ref}$),
$M_{\tilde{\chi}^{0}_{1}}= 86.4\ {\rm GeV/}c^{2}$, and 
$M_{\tilde{\tau}^{\pm}_{1}} = 152.7\ {\rm GeV/}c^{2}$.
Notice that $\tilde{\tau}^{\pm}_{1}$\ is the next to the lightest SUSY 
particle (NLSP) and decays  to $\tilde{\chi}^{0}_{1} \tau$
at 100\% of the time.
We assume in what follows that $M_{\tilde{\chi}^{0}_{1}}$\ and 
$M_{\tilde{\tau}^{\pm}_{1}}$\ 
will have been measured below the threshold of chargino pair production,
say at $\sqrt{s}$ = 310 GeV,
with uncertainties expected for
$\int {\mathcal L} dt = 100\ {\rm fb}^{-1}$\ \cite{stau2}.
In order to evaluate to what extent we can determine the chargino mass
in this setting, we then vary the $\Chione$\ mass around 171.0 GeV/$c^{2}$,
while keeping the $\Stau$ and $\Chizero$ masses unchanged, 
by appropriately adjusting the values of SUSY(-breaking) 
parameters~\footnote{For this we allow violation of the GUT relation 
between the U(1) 
and SU(2)$_{\rm L}$\  gaugino mass parameters (M$_1$ and M$_2$) 
that has been assumed 
for our reference point.}
and see how the observable distributions respond to the change. 

As for the ILC parameters, we assume
$\sqrt{s} = 400$\ GeV, $\int {\mathcal L} dt = 200\ {\rm fb}^{-1}$, and
an electron beam polarization of ${\mathcal P}(e^{-}) = -0.9$\ (left handed).
The $\tilde{\chi}^{+}_{1} \tilde{\chi}^{-}_{1}$\ production cross section
at ${\mathcal P}(e^{-}) = -0.9$\ is 20 times larger 
than that at ${\mathcal P}(e^{-}) = +0.9$\ (right handed). 
Furthermore, it can suppress the cross section for the
$e^{+}e^{-} \to \tilde{\tau}_1^+ \tilde{\tau}_1^-$ process that
yields almost the same final states as our signal process.

\section{Event Selection}
In our setting, the chargino pair production results in a final
state consisting of a $\tau$-pair with large missing energy.
The five major backgrounds for the signal event
are considered:
(i) $\tilde{\tau}^{\pm}_{1}$-pair production:
        $\tilde{\tau}^{+}_{1}
	\tilde{\tau}^{-}_{1} \rightarrow \tau^{+}
         \tilde{\chi}^{0}_{1} + \tau^{-}\tilde{\chi}^{0}_{1}$, 
(ii) neutralino pair production:
	$\Chizero \tilde{\chi}^{0}_{2} \rightarrow \Chizero + 
        \tilde{\tau}_{1} \tau \rightarrow 
        2\Chizero + \tau^{+} \tau^{-}$,
(iii) diboson production:
	 $W^{+}W^{-} \rightarrow 
         \tau^{+}\nu_{\tau} + \tau^{-}\overline{\nu}_{\tau}$,
	$Z^{0}Z^{0} \rightarrow
	\tau^{+}\tau^{-} + \ell^{+}\ell^{-}$\ ($\ell = e, \mu, \tau $) or 
        $\tau^{+}\tau^{-}+ \overline{\nu}\nu$,
(iv)  single boson production: $eeZ^{0}$ and $\nu\nu Z^{0}$\ with $Z^{0} \rightarrow
         \ell^{+} \ell^{-}$\ or $\nu \overline{\nu}$, and
(v) two-photon $\tau$-pair production: 
         $e^{+}e^{-} \rightarrow e^{+}e^{-} \tau^{+}\tau^{-}$.
As for the neutralino pair production, 
the SU(2)$_{\rm L}$\  relation 
between $\Chitwo$\ and $\Chione$\ masses is assumed.
We also assume 
the $\Chitwo$\ mass is measured 
via other processes such as $\Chitwo 
 \Chitwo \rightarrow 4\tau +2\Chizero$.
We ignore the uncertainty in the $\Chitwo$\ mass measurement, because
the effect in $\Chione$\ mass measurement  is 
found to be negligible.~\footnote{
 The $\Chione$\ mass determination accuracy is estimated to be 
	$\pm 0.02$\ GeV/$c^{2}$ if the $\Chitwo$ mass
fluctuation is $\pm 1$\ GeV/$c^{2}$.}
Notice that the cross section times branching fraction for $W$-pair 
production 
followed by $W \to \ell\nu$ decays 
is, for instance, eight times higher than that of the signal at our SUSY 
reference point.
We had thus better avoid events with purely leptonic $\tau$ decays.
Notice also that the two $\tau$-jets~\footnote{
A group of particles coming from a $\tau$ decay will hereafter be called 
a $\tau$-jet even if it contains only one charged particle.}
in the SUSY events are likely
to be more acoplanar than those from the SM backgrounds.
These observations lead us to 
the following selection criteria to improve
signal-to-background ratio:
\begin{list}{}{}
\item[(a)] there should be no $e/\mu$\ candidates positively identified in the
detector simulator;
\item[(b)] jet clustering has to yield two jets with $5\ {\rm GeV} \le E_{\rm jet} \le 160\ {\rm GeV}$\ 
      for  $y_{\rm cut} \ge 0.0025$~\cite{JADE};
\item[(c)] $-Q_{i}\cdot J_{iz}/|\vec{J}_{i}|
	      \le 0.8$, where $Q_{i}$\ and $\vec{J}_{i}$\ are the charge and 
	      the momentum of $i$-th jet with $i$ = 1(2) corresponding to 
the higher (lower) energy jet;
\item[(d)] $M_{\rm jet} \le 3\ $GeV/$c^{2}$;
\item[(e)] the missing transverse momentum $P^{miss}_{\rm T} \ge 20\ $GeV/$c$;
\item[(f)] $\cos\theta(J_{1}, P_{\rm vis}) \le 0.9$\ or 
	      $\cos\theta(J_{2}, P_{\rm vis}) \ge -0.7$, where
	      $\cos\theta(J_{i}, P_{\rm vis}) \equiv \vec{J}_{i}\cdot
              \vec{P}_{\rm vis}/(|\vec{J_{i}}||\vec{P}_{\rm vis}|)$ and
	      $\vec{P}_{\rm vis}$\ is the visible momentum vector 
	      calculated with both tracker and calorimeter information;
\item[(g)] thrust $\le 0.98$;
\item[(h)] acoplanarity angle $\theta_A$ $\ge 30^{\circ}$.
\end{list}
Cut (f) effectively rejects the $WW$\ background, because the
$\tau$ lepton from $W \rightarrow \tau\nu$\ tends to go in the parent 
$W$-boson momentum direction. 

We tabulate the resultant selection efficiencies after 
each cut for the signal and the background processes
at the reference SUSY point in Table~\ref{tab:eventsel}.
Total selection efficiency ($\epsilon_{\rm tot}$) for the
$\tilde{\chi}^{+}_{1}\tilde{\chi}^{-}_{1}$\ signal at the reference point
is $\epsilon^{\rm ref}_{\rm tot} =\  (15.8\pm 0.1)\%$.
We see that the selection criteria are effective at
suppressing the SM backgrounds, yielding
a signal-to-background ratio of 1.7.
\begin{sidewaystable} 
\begin{center}
\caption{Cumulative selection efficiencies ($\epsilon$) for 
signal ($\tilde{\chi}^{+}_{1}\tilde{\chi}^{-}_{1}$) and various 
sources of backgrounds.
The cross section includes branching ratios of $W \rightarrow e\nu, \mu\nu, 
\tau\nu$\ for $W^{+}W^{-}$\ process and 
$Z^{0} \rightarrow ee, \mu\mu, \tau\tau, \nu\nu$ for $Z^{0}Z^{0}$, 
$eeZ^{0}$, and $\nu\nu Z^{0}$\ processes.}
\label{tab:eventsel}
\vspace{0.5cm}
\begin{tabular}{lrrrrrrrr}
\hline
process & $\tilde{\chi}^{+}_{1} \tilde{\chi}^{-}_{1}$ & 
$\tilde{\chi}^{0}_{1} \tilde{\chi}^{0}_{2}$ & 
$\tilde{\tau}^{+}_{1} \tilde{\tau}^{-}_{1}$ & $W^{+}W^{-}$ & $Z^{0}Z^{0}$ &
$e e Z^{0}$ & $\nu \nu Z^{0}$ & 
\multicolumn{1}{l}{$e^{+}e^{-} \rightarrow$} \\
&(ref.) &(ref.)  & (ref.) & & & & & 
$e^{+}e^{-}\tau^{+}\tau^{-}$   \\
$\sigma$ (fb) (${\mathcal P}(e^{-}) = -0.9$)& 322 & 78 & 44 & 2304 & 83.6 
&1876 
& 92.8 & 5.62$\times 10^{4}$ \\
\hline
(a)\ $e/\mu$\ removal & 0.403 & 0.424 & 0.398 & 0.0747 & 0.0623 & 
$3.7\times 10^{-2}$ & 0.0437 & $7.3\times 10^{-1}$\\

(b)\ N$_{\rm jet}$ = 2 & 0.351 & 0.308& 0.348 & 0.0392 & 0.0522 &
$1.8\times 10^{-2}$ & 0.0377 &  $2.6\times 10^{-2}$ \\

(c)\ Jet angle cut & 0.240 & 0.218& 0.251 & 0.0122 & 0.0255 & 
$6.9\times 10^{-3}$ & 0.0239 &  $2.4 \times 10^{-2}$ \\

(d)\ $M_{\rm jet} \leq 3$ GeV/$c^{2}$ & 0.220 & 0.187 & 0.238 & 0.0118 & 
0.0229 & $5.8\times 10^{-3}$ & 0.0232 &  $2.4 \times 10^{-2}$ \\

(e)\ $P^{miss}_{\rm T} \geq 20$ GeV/$c$ & 0.182 & 0.147 & 0.198 & 0.0085
& 0.0208 & $4.8\times 10^{-4}$ & 0.0178 & $1.9 \times 10^{-4}$ \\  

(f)\ $\cos\theta (J_{1},P_{\rm vis}) \leq 0.9$ or & & & & & & & \\
\ \ \ \ \ $\cos\theta (J_{2},P_{\rm vis}) \geq -0.7$ & 0.175 & 0.142 & 0.185 & 
0.0054 & 0.0200 & $4.7 \times 10^{-4}$ &
 0.0161 & $1.2 \times 10^{-4}$ \\

(g)\ Thrust $\leq 0.98$ & 0.173 & 0.141 & 0.172 & 0.0043
& 0.0200 & $4.1 \times 10^{-4}$ & 0.0160 &  $1.1 \times 10^{-4}$ \\

(h)\ Acoplanarity $\theta_{A} \geq 30^{\circ}$ & 0.158 &0.137 &  0.158 & 0.0031
& 0.0186& $9.1\times 10^{-5}$ & 0.0136 &  $2.8 \times 10^{-5}$ \\
\hline
$\epsilon_{\rm tot}$ &  0.158 & 0.137 & 0.158 &  0.0031 & 0.0186 &
$9.1\times 10^{-5}$ & 0.0136 & $2.8\times 10^{-5}$ \\
(generated events) & 0.4M & 0.4M & 0.6M & 1M & 1M &
1.4M & 1M & 8M \\
\hline
\# of expected events & & & & & & & \\
for $\int {\mathcal L} dt$ = 200 fb$^{-1}$ 
& 10175 & 2137 & 1390  & 1428  & 311 & 34 & 252 & 315  \\
\hline
\multicolumn{8}{l}{$M_{\tilde{\chi}^{\pm}_{1}}=171.0\ {\rm GeV}/c^{2},
M_{\tilde{\tau}^{\pm}_{1}}=152.7\ {\rm GeV}/c^{2}$} \\
\end{tabular}
\end{center}
\end{sidewaystable}
As for the $ZZ \rightarrow \nu\nu q \bar{q}$\ event where quark jets are
mis-identified as $\tau$-jets, the total selection efficiency is 
small (less than $5.0\times 10^{-5}$) and the expected number of the
$ZZ \rightarrow \nu\nu q \bar{q}$\ events is 2.5. We do not include this
process in Table~\ref{tab:eventsel}.
We also show the signal selection efficiencies
for different chargino masses in Table~\ref{tab:eventsel1}.
The $\Stau$ and $\Chizero$ masses
are fixed to be 152.7 GeV/$c^{2}$\ and 86.4 GeV/$c^{2}$, respectively. 
\begin{table}
\begin{center}
\caption{Total selection efficiencies for five different chargino masses where
$M_{\tilde{\tau}^{\pm}_{1}}$ = 152.7 GeV/$c^{2}$\ and 
$M_{\tilde{\chi}^{0}_{1}}$
= 86.4 GeV/$c^{2}$.}
\label{tab:eventsel1}
\vspace{0.5cm}
\begin{tabular}{lrrrrr}
\hline
$M_{\tilde{\chi}^{\pm}_{1}}$ (GeV/$c^{2}$) & 161.0 & 166.0 & 171.0 &
176.0 & 181.0 \\
& & & (ref.) & & \\
$\Delta M (\equiv M_{\tilde{\chi}^{\pm}_{1}} - M_{\tilde{\tau}^{\pm}_{1}})$\ 
(GeV/$c^{2}$) & 8.3 & 13.3 & 18.3 & 23.3 & 28.3 \\ 
\hline
$\sigma$ (fb) & 433 & 378 & 322 & 268 & 214 \\
$\epsilon_{\rm tot}$ & 0.156 & 0.157 & 0.158 & 0.158 & 0.158 \\
\hline
\multicolumn{6}{l}{$\mu = 400$\ GeV, $\tan\beta = 50$, $\tau$\ polarization 
= 0.026, ${\mathcal  P}(e^{-}) = -0.9$} \\
\end{tabular}
\end{center}
\end{table}
The statistical uncertainty in the efficiency for each point is 
better than 0.1\%.
The efficiency depends on the $\Chione$\ mass only very weakly as long as
$\Delta M \equiv M_{\Chione} - M_{\Stau}$\ is large enough to deposit a jet
energy of $E_{\rm jet} \gtrsim $ 5 GeV in the detector.

\section{Results}
\subsection{Determination of chargino mass}
\begin{figure}
\begin{center}
\scalebox{0.48}{\includegraphics*{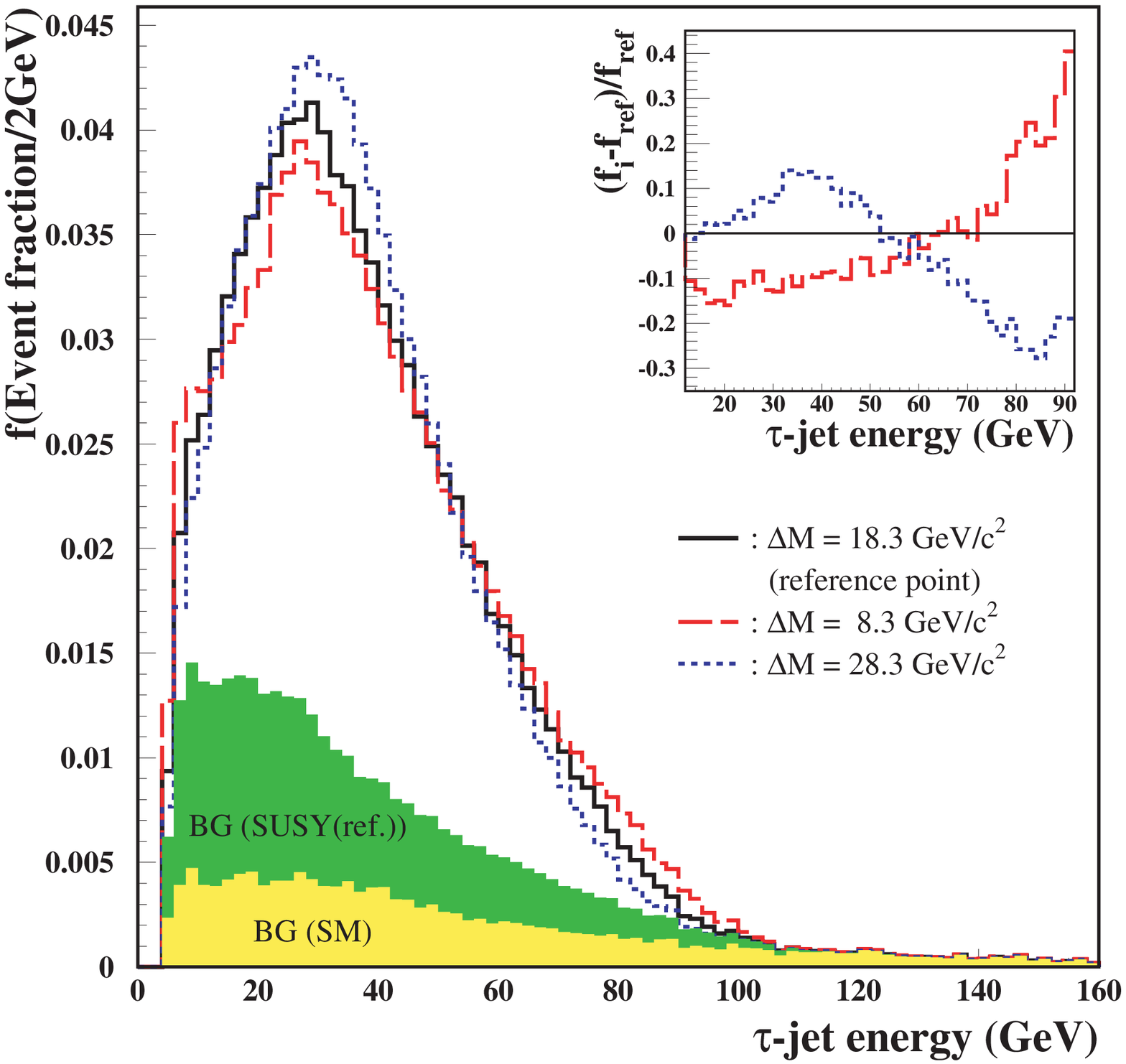}}
\caption{ Energy distributions of $\tau$-jets in the signal +  
backgrounds events for 3 representative choices of $\Delta M$\ after
our final selection criteria (Table~\ref{tab:eventsel}).
We also show background events. 
The shape in the reference SUSY scenario ($\Delta M = 18.3$\ GeV/$c^{2}$)
is compared to other cases ($\Delta M = 8.3$\ and 28.3 GeV/$c^{2}$). }
\label{fig:final2}
\end{center}
\end{figure}

If $\Chizero$\ and $\Stau$ masses are  predetermined, 
the $\tau$-jet energy distribution can be used
to determine the $\Chione$\ mass, since
the shape of the distribution, especially around the endpoint,
depends on $\Delta M$.
As $\Delta M$ increases, the peak of the distribution shifts to 
the higher energy side, while the endpoint moves to the lower
energy side.   
This can be seen in
Figure~\ref{fig:final2}, which plots the $\tau$-jet energy distributions 
for various $\Delta M$\ values. 

In order to determine the chargino mass using the $\tau$-jet
energy distribution, we first prepare a parametrized $\tau$-jet energy distribution
(a template) as a function of $M_{\Chione}$ in the following manner:
(i) the $\tau$-jet energy distribution, including the backgrounds, is fit 
to a polynomial function by MINUIT~\cite{MINUIT} for each of high statistics 
samples with different $\Chione$\ masses;
(ii) the fitting parameters are then parametrized as a function of 
$M_{\Chione}$. 
Although the $\tau$-jet energy distribution may depend 
on the remaining free parameters such as
$\tilde{\nu}$\ mass, $\mu$\ parameter, and $\tau$\ polarization, 
those are fixed in the present study. 

\begin{figure}
\begin{center}
\scalebox{0.48}{\includegraphics*{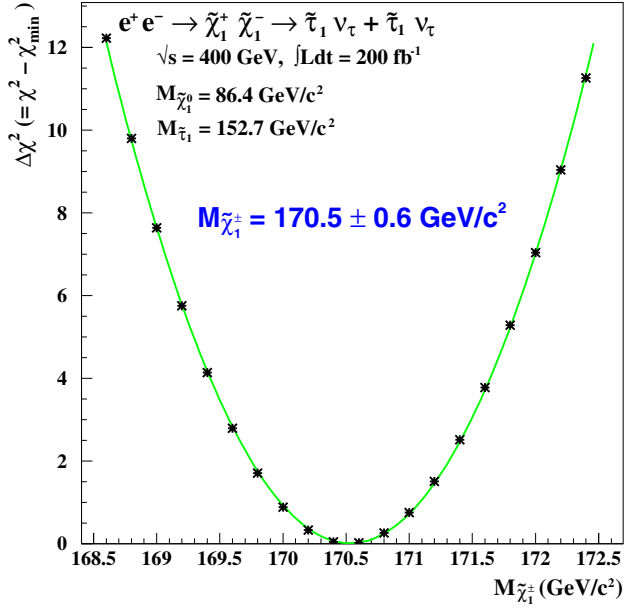}}
\caption{$\Delta\chi^{2}$ of $\tau$-jet energy distributions for
the reference point}
\label{fig:delW1SS}
\end{center}
\end{figure}

We then compare the $\tau$-jet energy distribution for a Monte Carlo sample
corresponding to 200 fb$^{-1}$\ with the template and calculate $\chi^{2}$.  
Figure~\ref{fig:delW1SS} shows $\Delta\chi^{2} \equiv \chi^{2} - 
\chi^{2}_{\rm min}$\ for the reference point 
($M_{\Chione}^{\rm ref}$\ = 171.0 GeV/$c^{2}$).
The fitted value of the $\Chione$\ mass is $170.5\pm 0.6$\ GeV/$c^{2}$.
This is about 1$\sigma$\ off the reference mass of 171.0 GeV/$c^{2}$.
We prepare approximately 100 statistically independent samples
of $\Chione \Chione$\ and $\Chitwo \Chizero$\ events for the 
reference point. 
Each sample is also corresponding to 200 fb$^{-1}$. The $\chi^{2}$\
fitting procedure provides consistently a $1 \sigma$\ uncertainty of 
$\pm 0.6$\ GeV/$c^{2}$\ for each sample. 
However the fluctuation of $\Chione$\ mass at the 
$\chi^{2}_{\rm min}$\ is $\pm 0.4$\ GeV/$c^{2}$\ around the 
$M_{\Chione}^{\rm ref}$. Therefore, we assign the total uncertainty in the
$\Chione$\ mass measurement to be $\pm 0.7$\ GeV/$c^{2}$\ (quadratic sum
of two sources). 

\begin{figure}
\begin{center}
\scalebox{0.48}{\includegraphics*{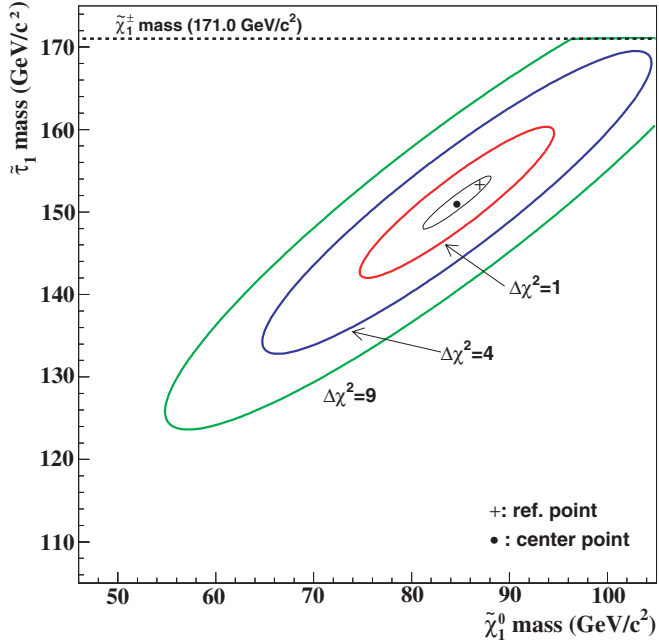}}
\caption{Contour plot in the $M_{\Stau} - M_{\Chizero}$\ plane, where
$M_{\Chione}$\ can be measured to an accuracy of
$\pm 0.7\ {\rm GeV/}c^{2}\ (\Delta\chi^{2} = 1)$, 
$\pm 1.4\ {\rm GeV/}c^{2}\ (\Delta\chi^{2} = 4)$, and 
$\pm 2.1\ {\rm GeV/}c^{2}\ (\Delta\chi^{2} = 9)$.
The smallest ellipse is the expected $1 \sigma$\  bound for
the $\Chizero$\ and $\Stau$\ mass fit with
100~fb$^{-1}$\ cited from Ref.~\cite{stau2}.}
\label{fig:contchi}
\end{center}
\end{figure}

The above estimate ignores the uncertainties in the input $\Stau$\ and 
$\Chizero$\ masses.
These uncertainties will affect the accuracy in 
the $\Chione$\ mass measurement, because the differences between 
$\Stau$, $\Chizero$, and $\Chione$ 
masses determine the shape of $\tau$-jet energy distribution. 
The higher endpoint, in particular, depends on the difference ($\Delta M$) between 
$\Stau$\ and $\Chione$\ masses.
In order to evaluate the impact of the uncertainties on the $\Chione$ mass
determination,
we repeat the study for $M_{\Chione} = 171.0\ {\rm GeV/}c^{2}$,
varying $M_{\Stau}$\ and $M_{\Chizero}$.
Figure~\ref{fig:contchi} shows a contour plot of 
$\Delta\chi^{2}$ in the $M_{\Stau}$\ - $M_{\Chizero}$\ plane. 
The center point of each $\Delta \chi^{2}$\ contour is at 
$M_{\Chizero}=84.6\ {\rm GeV/}c^{2}$\ and 
$M_{\Stau}=151.2\ {\rm GeV/}c^{2}$.
The contours clearly demonstrate a strong positive correlation
between $M_{\Chizero}$ and $M_{\Stau}$, implying that
the shape of the $\tau$-jet energy distribution
is mainly controlled by their difference, as long as the
$\Chione$ mass is fixed.
It is worth noting that the smallest ellipse, which is
the expected 1$\sigma$ bound from the $\Stau$-pair production study
with 100~fb$^{-1}$ cited from Ref.~\cite{stau2},
has a quite similar correlation and is well contained in the
contour corresponding to $\Delta \chi^2=1$.
The aforementioned expected accuracy
of $\pm 0.7$~GeV/$c^2$ for the $\Chione$ mass will hence essentially be
unaffected by the inclusion of the uncertainties in
the $\Chizero$ and $\Stau$ masses from the lower energy
$\Stau$-pair study.

\subsection{Production cross section measurement}
The production cross section times branching ratio is 
experimentally estimated by
$\sigma \cdot {\rm Br} = ({\rm N_{obs}} - {\rm N_{bg}})/
(\epsilon_{\rm tot} \cdot \int \mathcal{L} dt)$, 
where ${\rm N_{obs}}$\ is the number of observed events, ${\rm N_{bg}}$\ 
the number of expected background events, $\int \mathcal{L} dt$ the
integrated luminosity, and $\epsilon_{\rm tot}$\ the selection efficiency.
It should be noted that $\epsilon_{\rm tot}$\ weakly depends on 
$M_{\Chione}$\ as shown in Table~\ref{tab:eventsel1}: the efficiency varies by only 
0.6\% over the mass range of $\pm 20$\  GeV/$c^{2}$ 
from $M_{\Chione}^{\rm ref}$. 
This implies the possibility of extracting the cross section
times branching fraction ($\sigma \cdot {\rm Br} = 322 \pm 6$\ fb)
independently of the input $\Chione$\  mass, provided that 
$\int \mathcal{L} dt = 200 \pm 2$\ fb$^{-1}$
(the luminosity is assumed to be measured with an 
uncertainty of 1\%~\cite{stau2}),
${\rm N_{obs} - N_{bg} = 10175}$\ events (the background contribution is
fully estimated), and $\epsilon_{\rm tot} = 0.158 \pm 0.002$\ 
(the uncertainty includes the weak dependence of $\epsilon_{\rm tot}$\ 
on $M_{\Chione}$).
This indicates that we can determine the production cross section 
times branching
fraction to an accuracy of 2\% even without any precise $M_{\Chione}$\ study.

\section{Summary}
At large $\tan\beta$ values, the $\Stau$\ mass might be lighter than 
the $\Chione$\ mass within the MSSM framework.
In this case, the $\Chione \rightarrow \Stau \nu_{\tau}$\ decay will be 
dominant and that will be followed by $\Stau \rightarrow \tau^{\pm} \Chizero$.
We have thus studied the final state consisting of $\tau\tau$\ + 
large missing energy as a possible signature of the chargino pair 
production at a future $e^{+}e^{-}$\ 
linear collider. 

We proposed to use the $\tau$-jet energy distribution
to measure the $\Chione$\ mass in the large $\tan\beta$\ scenario.
Since this scenario leads $\Chizero$\ and $\Stau$\ masses to be lighter
than $\Chione$\ mass and opens the $\Stau$\ pair production mode with 
lower energy of $\Chione$\ pair production, the $\Chizero$\ and $\Stau$\ 
masses are likely to be determined in the lower $\Chione$\ production
threshold energy operation of the collider. 
The measurements of 171~GeV/$c^2$\ chargino could
then be made with an accuracy better than 1~GeV/$c^2$,
provided that the $\Chizero$\ and $\Stau$\ masses
will have been determined through
a $\Stau$-pair production study with 100~fb$^{-1}$.

Since the event selection efficiency does not significantly 
depend on the $\Chione$\ mass, the uncertainty in the production 
cross section shall be dominated by uncertainties in the selection 
efficiency and the luminosity measurement.
We expect to determine the cross section to an accuracy of 2\% at the
reference point despite the cascade decay mode.

\ack
The authors would like to thank R.~Arnowitt for carefully reading the 
manuscript.
Y.K. is supported by the Grant for Kinki University, while
K.F. is supported in part by Japan-Europe (UK) Research Cooperation 
Program and JSPS-CAS Scientific Cooperation Program under the 
Core University System, T.K. and V.K. by U.S. Department of Energy grant 
DE-FG02-95ER40917, and M.M.N. by the Grant-in-Aid for Science Research, 
Ministry of Education, Science and Culture, Japan 
(No.14540260 and No.14046210), respectively.

\end{document}